\documentclass[9pt,twocolumn,twoside]{article}
\usepackage{array}
\usepackage{rotating}
\usepackage{multirow}
\usepackage{mwe,tikz}\usepackage[percent]{overpic}
\usepackage{color}
\usepackage[colorinlistoftodos]{todonotes}
\usepackage[normalem]{ulem}
\usepackage[affil-it]{authblk}
\usepackage{amsmath}

\title{Spectrally resolved single-shot wavefront sensing of broadband high-harmonic sources}

\author[1,2,$\ddagger$]{L. Freisem}
\author[1,2,$\ddagger$]{G. S. M. Jansen}
\author[1,2]{D. Rudolf}
\author[1,2]{K. S. E. Eikema}
\author[1,2,*]{S. Witte}

\affil[1]{\small Advanced  Research  Center  for  Nanolithography  (ARCNL), Science  Park  110,  1098  XG  Amsterdam, Netherlands} \affil[2]{Dept.  of  Physics  and  Astronomy, Vrije  Universiteit, De  Boelelaan  1081,  1081  HV  Amsterdam, Netherlands\newline}

\affil[$\ddagger$]{These authors contributed equally to this work}
\affil[*]{email: witte@arcnl.nl}

\date{}
\begin{document}



\twocolumn[
\begin{@twocolumnfalse}
\maketitle

\begin{abstract}
\noindent \normalsize Wavefront sensors are an important tool to characterize coherent beams of extreme ultraviolet radiation. However, conventional Hartmann-type sensors do not allow for independent wavefront characterization of different spectral components that may be present in a beam, which limits their applicability for intrinsically broadband high-harmonic generation (HHG) sources. 
Here we introduce a wavefront sensor that measures the wavefronts of all the harmonics in a HHG beam in a single camera exposure. By replacing the mask apertures with transmission gratings at different orientations, we simultaneously detect harmonic wavefronts and spectra, and obtain sensitivity to spatiotemporal structure such as pulse front tilt as well. 
We demonstrate the capabilities of the sensor through a parallel measurement of the wavefronts of 9 harmonics in a wavelength range between 25 and 49~nm, with up to $\lambda/32$ precision.
\end{abstract}
\vspace{1cm}

\end{@twocolumnfalse}
]



\section{Introduction}
Accurate measurements in imaging and optical metrology often rely on a precise knowledge of the incident beam parameters. The Hartmann wavefront sensor, first introduced in the year 1900 to calibrate telescopes \cite{Hartmann}, quickly became a standard tool to characterize optical wavefronts. The sensor consists of an opaque plate containing a structured array of apertures, and the wavefront information is retrieved by measuring the propagation of light transmitted through the holes. 
Detecting the positions of the aperture transmission with respect to a known reference  provides amplitude and local phase tilts of a monochromatic wavefront simultaneously, from which the wavefront is then reconstructed. A widely used modification to increase the sensitivity of Hartmann sensors is the lenslet array introduced by Platt and Shack \cite{Shack} in 1971.
In the regime of extreme ultraviolet (XUV) radiation, where lens arrays are challenging, the traditional hole array was shown to be a useful approach for the characterization of monochromatic synchrotron beams~\cite{mercere_hartmann_2003} and high-harmonic generation (HHG) sources~\cite{valentin_high-order_2008}.

\begin{figure*}[!th]
\begin{centering}
\includegraphics*[width=\textwidth]{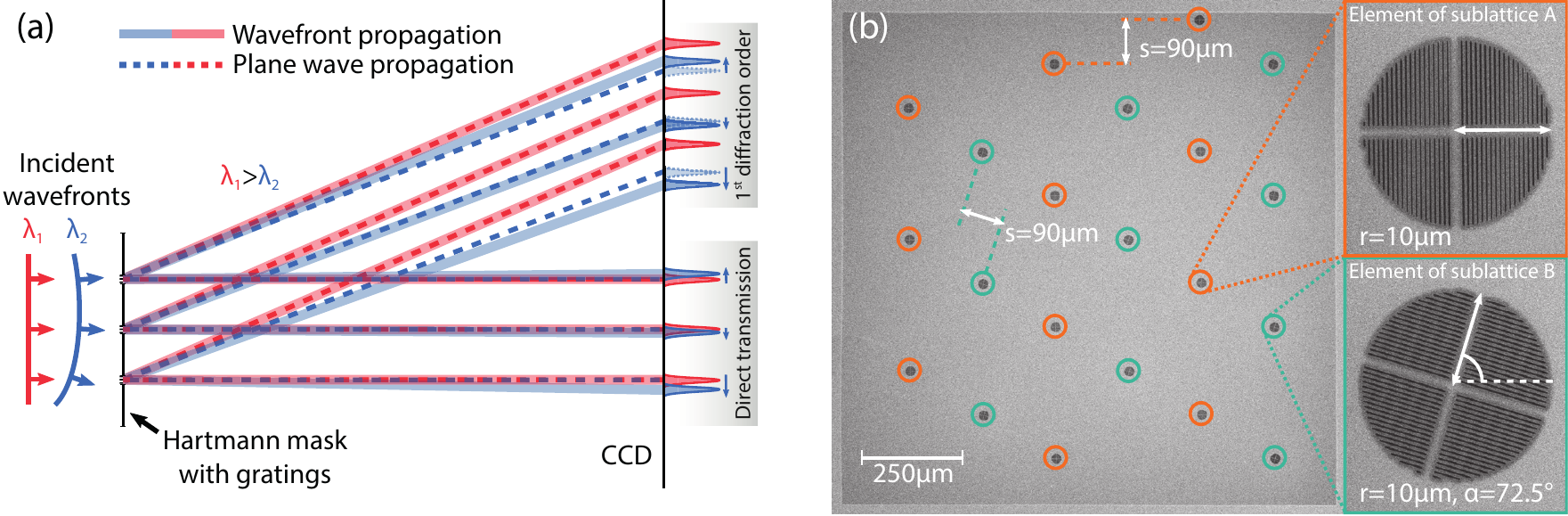}
\par\end{centering}
\caption{\label{fig:Scheme}\textbf{(a)} Diffraction scheme of the multi-wavelength Hartmann sensor. The direct transmission spots of all spectral components overlap and cannot be distinguished. In contrast, in the first diffraction order the different spectral components are spatially separated, while still containing the wavefront information through the displacement of the spots with respect to a set of reference positions. \textbf{(b)} Scanning electron microscope image of the sample, with two insets showing magnified examples of the two main grating alignment groups. The gratings of one subset are rotated by an angle $\alpha = 72.5^\circ$, ensuring that the diffraction does not overlap with the diffraction of the other grating group. To improve the structural stability of the $d=0.5~\mu\mathrm{m}$ pitch gratings, additional $1~\mu\mathrm{m}$-wide support bars are used, forming a cross through their center.}
\end{figure*}

High-harmonic generation sources \cite{popmintchev_attosecond_2010,bartels_generation_2002,rudawski_high-flux_2013} are becoming a mature table-top source of coherent XUV radiation, used in many applications such as nanoscale imaging~\cite{gardner_subwavelength_2017,zurch_real-time_2014,baksh_wide-field_2016}, soft-X-ray spectroscopy~\cite{cousin_high-flux_2014} and attosecond physics~\cite{krausz_attosecond_2009,kraus_measurement_2015,silva_spatiotemporal_2015}. 
However, a significant limitation of Hartmann sensors is their inability to provide spectral sensitivity. This is particularly important for the characterization of HHG beams, as it is known that the wavefronts of different harmonics can be substantially different~\cite{zerne_phase-locked_1997}. Spectral wavefront variations in HHG may arise as a result of the phase matching geometry~\cite{He_2009,frumker_order-dependent_2012}, and can convey information about quantum path interferences in the HHG process~\cite{schapper_spatial_2010}. 
Introducing a controlled spectrally dependent wavefront tilt actually is the basis for the attosecond lighthouse effect, which can be used to produce isolated attosecond pulses~\cite{quere_attosecond_2012,kim_photonic_2013}.
To extract spectrally resolved wavefront information, several methods have been demonstrated, such as slit scanning combined with grating-based spectrometry~\cite{frumker_frequency-resolved_2009,lloyd_complete_2013}, and lateral shearing interferometry~\cite{austin_lateral_2011,mang_simultaneous_2014}. While these methods retrieve spectrally resolved wavefront information, they all depend on some form of mechanical scanning in obtaining a complete dataset, and therefore can only measure average wavefronts over many pulses. Due to the nonlinear nature of the HHG process, the influence of small driving pulse variations on the phase matching process can lead to wavefront fluctations on a shot-by-shot basis. Such fluctuations may in turn have consequences in experiments where knowledge of the input parameters typically limits the achievable accuracy, such as EUV scatterometry~\cite{ku_euv_2016} and coherent diffractive imaging~\cite{ge_impact_2013}.

In this article we introduce a novel wavefront sensor concept called a Spectroscopic Hartmann Mask (SHM), that provides wavefront data for multiple spectral components directly in a single camera exposure.
We replace the apertures of the Hartmann mask by transmission gratings which, in addition to the regular Hartmann spot pattern, produce replicas of this spot pattern at the $\pm 1^{\mathrm{st}}$~diffraction orders as shown schematically in Fig.~\ref{fig:Scheme}(a). These diffracted beams have similar sensitivity to local wavefront tilts at each aperture, while the different harmonics can be clearly separated in the image. 
This approach enables the isolation and analysis of Hartmann spot patterns for the individual wavelengths of all harmonics in a HHG beam simultaneously from a single camera recording. The SHM wavefront sensor concept enables fast characterization of full HHG beams, even on a single-pulse basis for sufficiently bright beams, providing a unique tool for characterization of broadband short-wavelength sources and sensing of the underlying physics of the HHG process.

\begin{figure*}[!tb]
\begin{centering}
\includegraphics[width=0.65\textwidth]{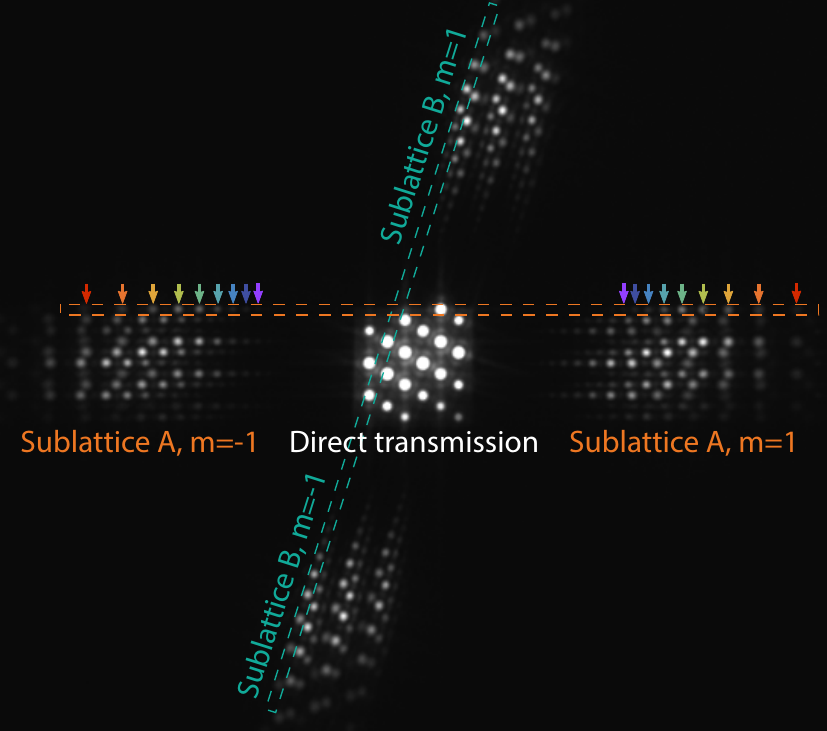}
\par\end{centering}
\caption{\label{fig:Diff}Diffraction pattern of the spectroscopic Hartmann mask, obtained using high harmonics generated in argon. This image was taken using a one second exposure. In addition to the central group of spots from direct transmission, there are four groups of diffraction spots, which correspond to the $\pm 1^{\mathrm{st}}$ diffraction orders of the different gratings. 
All these spots are distributed along the two main diffraction directions, corresponding to the grating orientations as pointed out in Fig. \ref{fig:Scheme}(b). As a guide to the eye, the diffraction spots corresponding to one horizontal and one vertical grating are highlighted by orange and green boxes. For all apertures, the individual harmonics are spatially separated in the first diffraction orders, enabling wavefront reconstruction.}
\end{figure*}

\section{Spectrally resolved wavefront characterization}
\subsection{Spectroscopic Hartmann mask design}
A scanning electron microscope image of the SHM is displayed in Fig.~\ref{fig:Scheme}b, along with insets that show the individual gratings in more detail.
To ensure clear separation of the spot patterns for the different harmonics, the mask design and the distance to the camera are important parameters. If the SHM-to-camera distance is too small, the first-order diffraction spots are not spread out sufficiently, leading to overlapping signal on the detector. Conversely, if the distance is too large, diffraction resulting from the apertures leads to larger spot sizes, resulting in overlapping spots of neighboring apertures. Using Fresnel diffraction theory to simulate the detected intensity patterns, it is readily possible to find designs that satisfy the requirements for different HHG spectra. The SHM design shown in Fig.~\ref{fig:Scheme}b enables wavefront measurements for all harmonics produced in argon using a fundamental wavelength of 810~nm. Moreover, there is some flexibility for measurements at both longer and shorter wavelengths by changing the distance between mask and camera. 

Compared to conventional Hartmann masks, the SHM design does have a significantly lower spatial density of apertures. This requirement stems from the fact that each aperture now leads to a multitude of diffraction spots for the different harmonics, which all need to be spatially separated in order to retrieve accurate wavefront data. However, the sampling density in the presented design still allows wavefront characterization up to the fourth-order Zernike polynomials, which is sufficient for many practical purposes. Optimized designs with increased sampling density can be envisaged with more modelling efforts.

The SHM is fabricated by milling holes into a 200~nm thick, $1\times1$~mm sized silicon nitride membrane, covered with 5~nm chromium and 27~nm gold, using the focused ion-beam technique. The individual apertures are circular with a radius $r=10~\mu$m, and contain a transmission grating with a pitch $d=0.5~\mu$m. With this design, positioning the SHM at 3~cm from the camera ensures clear separation of high harmonics in the 20-60~nm wavelength range generated using 810~nm fundamental wavelength, while diffraction from the circular apertures themselves remains limited. 
To avoid overlap between the diffraction from adjacent apertures, the gratings are tilted at $17.5^{\circ}$ with respect to the row of apertures, which ensures that the rows of diffraction spots are separated by 90~$\mu$m on the camera. In addition, we orient the gratings in two different directions. Having multiple diffraction directions allows a higher spatial sampling density of the beam while avoiding overlapping spots on the camera. Furthermore, it enables detection of systematic effects (more detail given below), reducing the sensitivity to angular misalignment and making the sensor essentially self-referencing.

\subsection{Measuring HHG wavefront data}
To demonstrate spectrally resolved wavefront measurements of high-harmonic beams, we use the output of a Ti:sapphire-seeded non-collinear optical parametric chirped-pulse amplifier (NOPCPA). The 1~mJ, 25~fs, 810~nm laser pulses are focused with an $f=25$~cm lens into a pulsed gas jet, resulting in harmonics produced in argon or krypton covering the $25-56$~nm wavelength range. We use an adjustable iris to optimize HHG phase matching, leading to a Gaussian EUV beam. A $200$~nm thick aluminum foil is used to block the infrared radiation. An XUV-sensitive CCD-camera (Andor Ikon-L) is used to detect the radiation at approximately 63~cm away from the HHG source. The SHM is mounted on a translation stage and positioned in the beam at 3~cm before the camera. 

A camera recording of the SHM diffraction pattern is shown in Fig.~\ref{fig:Diff}. Each aperture leads to an array of diffraction spots in the direction perpendicular to the grating lines, corresponding to the individual harmonics. As a guide to the eye, the diffraction from two apertures with different grating orientations are highlighted by the dashed boxes (note that the color coding corresponds to Fig.~\ref{fig:Scheme}(b)). 
In addition to the normal Hartmann spot pattern from the direct transmission in the center of the image, the individual harmonics produce clearly separated diffraction spots in the $\pm 1^{\mathrm{st}}$ diffraction orders, of which the positions can be determined accurately by a standard two-dimensional Gaussian fitting procedure. As schematically depicted already in Fig.~\ref{fig:Scheme}(a), these first-order diffraction spots are sensitive to the local wavefront tilts. Therefore they can be used to do wavefront reconstruction, for all harmonic orders simultaneously. In the present experiment up to nine harmonics are observed: the wavefronts of all the individual harmonics can be reconstructed from the single camera image of Fig.~\ref{fig:Diff}. To increase the accuracy of the spot determination further, we recorded images with exposure times up to 5~seconds in argon, and 10~seconds in krypton for the results described in section~\ref{sec:results}.

\begin{figure}
\begin{centering}
\includegraphics[width=\linewidth]{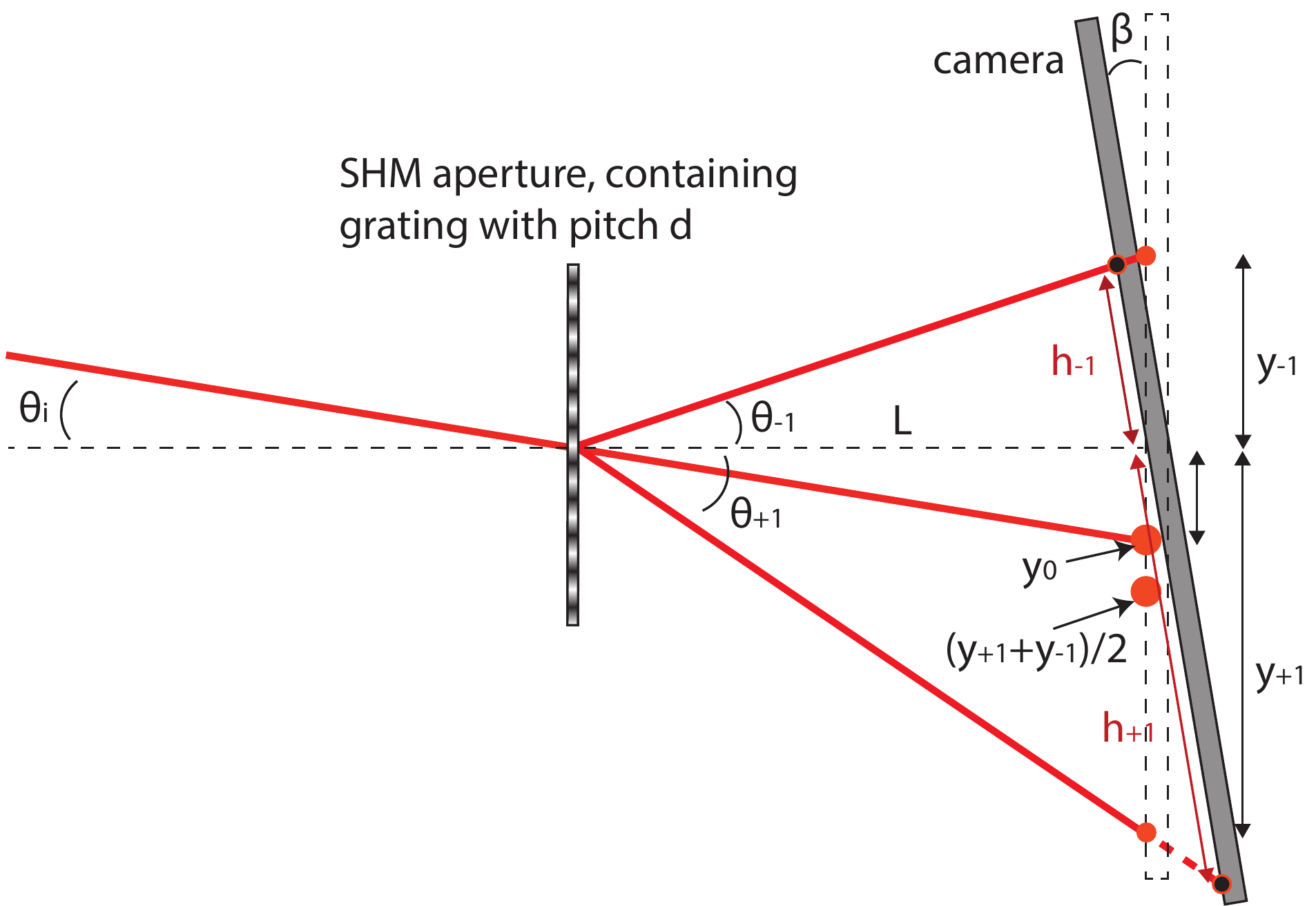}
\par\end{centering}
\caption{\label{fig:angle_scheme} 
Schematic overview of the transmission of a single spectroscopic Hartmann aperture. The center of mass of the +1 and -1 diffraction orders approximates the position of the direct transmission. The quality of this approximation depends on the incident wavefront tilt as well as the angle between camera and wavefront sensor.}
\end{figure}

\subsection{SHM diffraction pattern analysis}
Once the respective positions of all diffraction spots corresponding to each wavelength have been determined, all the required information for reconstructing the wavefronts is in principle available. Because of the diffraction geometry, there are some subtle differences compared to the traditional Hartmann mask analysis. A schematic overview of the diffraction geometry for a single aperture is drawn in Fig.~\ref{fig:angle_scheme}. At a distance $L$ behind the SHM, the $\pm 1^{\mathrm{st}}$ diffraction orders appear at positions $y_{\pm 1}$. The main principle of the SHM is that a nonzero wavefront tilt angle $\theta_i$ leads to a shift of the $\pm 1^{\mathrm{st}}$ diffraction orders that is similar to the shift in the direct transmission position $y_0$. Therefore, the center of mass $(y_{+1} + y_{-1})/2$ of these diffraction positions can be used as input for the wavefront reconstruction algorithm. However, as indicated in Fig.~\ref{fig:angle_scheme}, there are two additional effects that need to be taken into account. First of all, for a non-zero $\theta_{i}$, the shift of the +/-1 diffraction orders is not perfectly symmetric, leading to an offset between their combined center of mass position and $y_0$. Secondly, there may be a finite angle $\beta$ of the camera with respect to the SHM. In this case, the measured positions of the diffraction spots are given by $h_{\pm 1}$, which are related to the expected positions as $h_{\pm 1} = y_{\pm 1} \cos{\theta_{\pm 1}} / \cos{\left(\theta_{\pm 1} \mp \beta\right)}$. Overall, the position of the measured spot position for the $m^\mathrm{th}$ diffraction order at a wavelength $\lambda$ can be written as: 
\begin{equation}
h_m = \frac{L(m \lambda - d \sin(\theta_i))}{d \sin(\beta+\arccos(\frac{m \lambda}{d}-\sin(\theta_i)))}
\label{Eq:hm}
\end{equation}
where $d$ is the grating pitch and $L$ is the distance between SHM and camera.

Starting from Eq.~\ref{Eq:hm} and using a Taylor approximation to quantify the effect of small variations in $\theta_i$ and $\beta$, we find an expression for the center of mass position:
\begin{multline}
(h_{-1}+h_{1})/2 \approx \\ h_0 + \frac{L \lambda^2}{\lambda^2-d^2}\beta+(L-\frac{L d^3 \sqrt{d^2-\lambda^2}}{(d^2-\lambda^2)^2})\theta_i
+ O(\beta^2 \theta_i + \beta \theta_i^2).
\label{Eq:taylor_angles}
\end{multline}
From this expression, it is clear that the effects of the two angles are independent up to the third-order Taylor terms, so that these effects can readily be mitigated. 
As $\beta$ is a constant number for all apertures, the effect of the $\beta$-dependent term is to shift the center of mass position of all diffraction spots of a given wavelength by an equal amount. Therefore only the linear Zernike terms (tip and tilt) would be affected. By having multiple grating orientations in the SHM, the effect of this constant offset can be fully eliminated, as only the subset of apertures diffracting in the plane of rotation of $\beta$ is affected by this offset. The presence of an angle $\beta$ in either direction will therefore show up as a different tip/tilt measurement for both subsets of apertures, from which the offsets can be deduced and corrected for. As a result, spectrally-dependent variations in tip and tilt can be quantified with the SHM.

The $\theta_i$-term in Eq.~\ref{Eq:taylor_angles} means that the center of mass position displaces more than the direct transmission in case of a local wavefront tilt along the grating diffraction direction. This effect needs to be corrected for to avoid a slight overestimation of the wavefront tilt. The size of this error depends only on known parameters and can be calculated to be $\sim 0.5\%$ per degree wavefront tilt at a wavelength of 30~nm. It it therefore straightforward to correct for this effect by first performing a wavefront reconstruction, finding an initial value for $\theta_i$, and subsequently correcting the center of mass position using the term from Eq.~\ref{Eq:taylor_angles}. This correction could be performed in multiple iterations, but we find that one correction step suffices for an accurate wavefront reconstruction. Finally, the higher-order terms in Eq.~\ref{Eq:taylor_angles} are found to be at least a factor $10^3$ smaller than the linear terms and can be safely ignored. 


\subsection{Wavefront reconstruction}
The analysis of the SHM diffraction patterns yields a set of Hartmann spot patterns for all the individual harmonics. These patterns can then be compared to the reference structure to find a displacement for every aperture on the sensor. This comparison yields the gradient of the wavefront $\nabla\phi\left(\vec{r}\right) = \Delta \vec{r} /L$, in which $\Delta r$ is the spot displacement with respect to the calibration. We characterized the reference pattern at 0.7~$\mu$m spatial resolution from an image of the SHM recorded by an optical microscope. This pattern is then rotated and centered with respect to the measured data.

A calibration of the distance $L$ is required for the analysis. To this end we recorded the SHM pattern from Krypton harmonics with the sensor placed at three different distances from the camera, with a precisely known displacement $\Delta L$ between them. These images allowed calibration of the harmonic wavelengths from the measurement geometry (Fig.~\ref{fig:angle_scheme}), upon which a consistent value for $L$ could be retrieved. As an independent verification of the wavelengths of the different harmonics, spatially resolved Fourier-transform spectroscopy (FTS)~\cite{Jansen2016} was used to measure the spectra of the individual spots on the camera.  
In our case where multiple harmonics are detected, $L$ can in principle also be obtained from the spot separation of consecutive harmonics on the camera image, if the assumption of an equal frequency spacing of twice the driving laser frequency between harmonics holds. We find that this approach leads to a slightly less accurate wavelength calibration, but still provides good wavefront reconstructions from a single SHM pattern.

From the measured local wavefront tilts the total wavefront can then be reconstructed using a two-dimensional integration of the measured gradient. In order to do so, a bi-cubic spline interpolation is used to resample the data from the non-rectangular arrangement of subpupils to convert it to a rectangular grid. The complete wavefront is obtained by numerical integration of the gradient.

\begin{figure}[!t]
\begin{centering}
\includegraphics[width=0.9\linewidth]{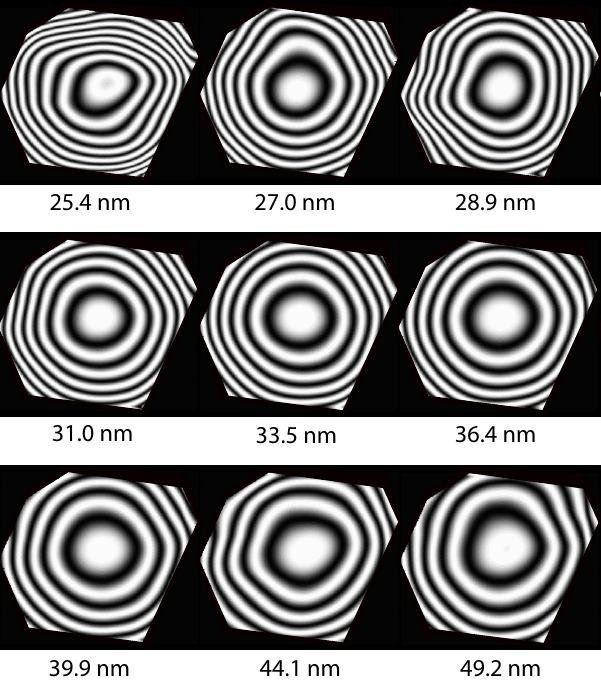}
\par\end{centering}
\caption{\label{fig:WFs} 
Reconstructed wavefronts for nine harmonics produced in argon, visualized by calculating their interference pattern with a plane wave. The center wavelength of the harmonic is indicated below each image.} 
\end{figure}

\section{Wavefront characterization of a high-harmonic beam}
\label{sec:results}
\subsection{Wavefront reconstructions of individual harmonics}
Figure~\ref{fig:WFs} shows the reconstructed wavefronts for a range of nine harmonics generated in argon. To visualize these wavefronts and show their small features in addition to the curvature, they are displayed as how they would interfere with a plane wave. As the HHG beam undergoes free-space propagation from its generation point to the SHM, we measure a diverging beam with a significant spherical wavefront curvature. The brightest harmonics (33.5, 36.4 and 39.9~nm) appear to be nearly diffraction-limited, while the other harmonics show some deviations from an ideal spherical wavefront. 

Because an uncertainty in the position of each aperture affects the final wavefront reconstruction in a different and complex way, we analyze the achieved wavefront accuracy by a Monte Carlo type analysis. From the center positions and uncertainties of the Gaussian fits of all the individual diffraction spots, we construct a distribution of Hartmann spot patterns that are all statistically consistent with the original measurement. We reconstruct wavefronts for all these spot patterns and compare the results. For the measurement shown in Fig.~\ref{fig:WFs}, we find a wavefront reproducibility better than $\lambda/9$. The result is more accurate for the brightest harmonics, which show an uncertainty of $\lambda/32$.

\begin{figure}[!ht]
\begin{centering}
\includegraphics*[width=0.92\columnwidth]{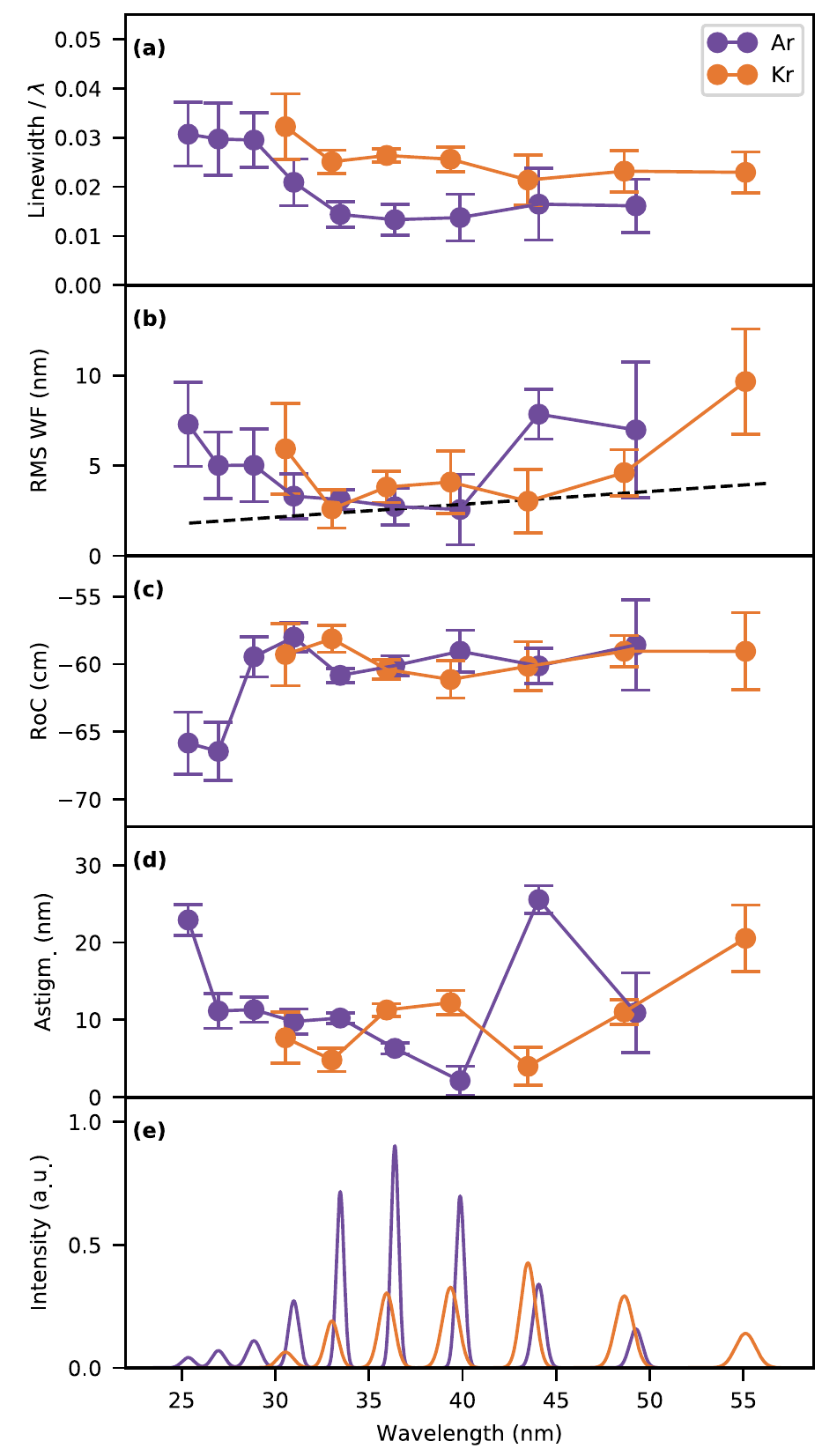}
\par\end{centering}
\caption{\label{fig:ZP} Wavelength-dependence of several beam parameters retrieved from a SHM measurement on HHG in Ar and Kr. Error bars constitute one standard deviation. \textbf{(a)} Linewidth of individual harmonics. \textbf{(b)} RMS deviation of the wavefronts from a spherical wave. The Marechal-criterion of $\lambda/14$ is indicated by a dashed line. 
\textbf{(c)} Radius of curvature (RoC) of the respective HHG wavelengths. \textbf{(d)} Astigmatism magnitude as calculated from the Zernike terms. \textbf{(e)} Reconstructed HHG spectra, normalized to the brightest argon line. The krypton spectrum is scaled by $2\times$ for visibility purposes.} 
\end{figure}

The information acquired from such a SHM measurement is not limited to wavefronts. Since each SHM aperture is a small transmission grating, a spectrum can be obtained for all apertures in the sensor. This enables a measurement of the linewidth of the harmonics as well as the spatial intensity distribution. Since a non-zero linewidth would result in elliptical spots, an accurate measurement of the linewidth can be made by fitting the spots with a 2D Gaussian that has different widths parallel and perpendicular to the diffraction direction. The width of each harmonic can then be extracted by deconvolution as $\mathrm{FWHM}=\sqrt{\left(\sigma_w^2-\sigma_h^2\right)}$, where $\sigma_w$ and $\sigma_h$ are the $1/e$-widths in the parallel and perpendicular directions respectively. This analysis assumes that the Gaussian shape is appropriate for both the spectrum and the diffraction spot shape.

\subsection{Spectroscopic wavefront analysis}
To quantify the wavefront aberrations we expand the retrieved wavefronts in terms of Zernike polynomials, including the lowest 11 polynomial terms from $Z_0$ to $Z_3$ and $Z_4^0$~\cite{wyant_basic_1992}. The unit circle on which the Zernike polynomials are defined has a diameter of 740~$\mu$m and is chosen to fit within the region covered by the sensor apertures, ensuring accurate Zernike coefficients within this region. 

Figure~\ref{fig:ZP} shows a set of results from such a wavefront analysis in terms of Zernike polynomials over a range of detected harmonic wavelengths, for HHG produced in both argon and krypton. As discussed above, the SHM provides some information about spectral parameters such as the linewidth of the harmonics, which is displayed in Fig.~\ref{fig:ZP}(a). At shorter wavelengths the harmonics are more broadband, and we observe broader harmonics in krypton than in argon. The harmonics in krypton also appear slightly blue-shifted (Fig.~\ref{fig:ZP}(e)), which together with the broader harmonics is indicative of the rather high intensity used for HHG in krypton. 
Figure~\ref{fig:ZP}(b) shows the root-mean-square (RMS) deviation of the wavefronts from a perfect spherical shape, which can be viewed as a measure of the deviations from a diffraction-limited wavefront. Within the measurement accuracy, the wavefront deviations are found to be either at, or slightly above, the Marechal criterion of $\lambda/14$ (dashed line), indicating that the XUV beams are in general close to diffraction-limited.

While the strongest harmonics appear close to diffraction-limited, stronger wavefront aberrations are observed for the weaker harmonics. As an example, Fig.~\ref{fig:ZP}(d) shows the magnitude of astigmatism, calculated as the root mean square of the $Z_2^{+2}$ and $Z_2^{-2}$ Zernike terms corresponding to straight and diagonal astigmatism. We found astigmatism to form the main contribution to the wavefront deviations observed in Fig. \ref{fig:WFs}, although there are small amounts of coma and trifoil present in the wavefronts as well.  

One further parameter which is of interest is the radius of curvature (RoC) of the spherical wavefront component of the individual harmonics, as it has previously been reported that there can be a variation of the apparent focal distance with high-harmonic number~\cite{frumker_order-dependent_2012}. Retrieving the RoC from the SHM data involves a combination of the defocus, spherical aberration and astigmatism Zernike terms~\cite{wyant_basic_1992}.  
From our measurement data, the observed radii of curvature (Fig.~\ref{fig:ZP}(c)) roughly match the source-to-mask distance of $60~\mathrm{cm}$, but no clear trend can be observed. This may be attributed to different phase-matching conditions in the high-harmonic generation. At the shortest wavelengths a deviation towards longer RoC is retrieved, although the error bars for these low-intensity harmonics also increase significantly.
\begin{figure}[!t]
\begin{centering}
\includegraphics*[width=\columnwidth]{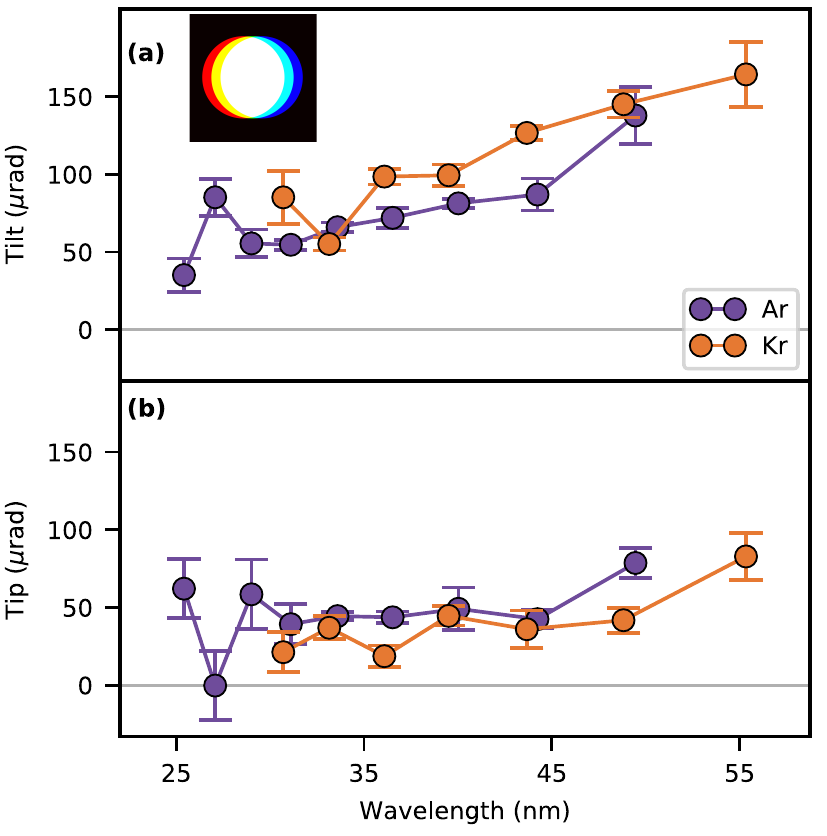}
\end{centering}
\caption{\label{fig:Misalign} Measured relative wavefront tilt in the horizontal (tilt, (a)) and vertical (tip, (b)) directions as a function of wavelength, retrieved from the SHM analysis, for HHG produced in argon and krypton with 810~nm fundamental wavelength. A significant wavelength-dependent wavefront tilt in the horizontal plane (tilt) is observed, indicating the presence of a spatial chirp of the HHG beam as schematically drawn in the inset to (a).}
\end{figure}

\subsection{Detection of HHG pulse front tilts}
A remarkable feature of the SHM is the ability to measure relative wavefront tilts between the individual harmonics. Especially for ultrafast pulsed sources, the ability to measure tilts allows a characterization of spatiotemporal coupling such as pulse front tilts, and associated effects such as the attosecond lighthouse effect~\cite{quere_attosecond_2012,kim_photonic_2013}. Since a wavefront tilt leads to a constant displacement of all spots on the camera, an accurate tilt measurement can only be performed with a Hartmann sensor if the exact position of the reference grid with respect to the camera is calibrated. For the SHM, an absolute wavefront tilt can in principle be determined if an accurate calibration of $\beta$ is performed. Such a calibration would require measurements of at least two collinear monochromatic beams with different wavelengths and is challenging to perform, while the physical significance of an absolute tilt measurement is limited. 


Knowledge of a possible relative tilt between different harmonics is relevant for many ultrafast experiments, as it constitutes a spatial chirp, which is equivalent to a tilt of the pulse front. Such a spatial chirp can modify the outcome of experiments that are sensitive to space-time coupling of ultrashort pulses. 
In a SHM wavefront measurement, we can use any of the detected harmonics as reference. 
Because in a SHG measurement the wavefronts of all harmonics are measured simultaneously in a common geometry,  the relative difference in tip and tilt for different harmonics can readily be measured. 

In this analysis the effect of a possible camera tilt needs to be included. As already discussed before with Eq.~\ref{Eq:taylor_angles}, a camera tilt along the diffraction direction leads to an additional displacement of the center-of-mass positions that needs to be corrected to obtain correct spot positions for wavefront reconstruction.
Since the displacement perpendicular to the grating direction is only sensitive to the wavefront tilt, an SHM containing gratings with two different directions can directly isolate this effect and measure the actual wavefront tilt relative to the reference wavelength. 

In this way we extract relative wavefront tilts for the individual high-harmonics produced in argon and krypton. The resulting wavefront tilt and tip are shown in Figs.~\ref{fig:Misalign}(a) and~\ref{fig:Misalign}(b). We find wavefront-tilt variations up to 0.1 milliradians across our full HHG spectrum. In comparison, the divergence of the high harmonic beams is approximately 1 milliradians, which means that even though the individual harmonics remain largely overlapped, there is a spectral variation and associated pulse front tilt in the HHG beam. We observe similar pulse front tilts in argon and krypton, indicating that this tilt is most likely due to a pulse front tilt in the driving laser beam ~\cite{quere_attosecond_2012,kim_photonic_2013}. This is to be expected, as the employed beam path contains some birefringent components with a slight intentional misalignment in the horizontal direction~\cite{Jansen2016}. Similar geometries have been used to intentionally induce strong tilts in the HHG beam, leading to the attosecond lighthouse effect~\cite{kim_photonic_2013}. The SHM provides a sensitive measurement device to quantify the presence of even small tilts, aiding the alignment and optimization of broadband HHG wavefronts and spatiotemporal couplings.

\section{Conclusions}
In summary, we have demonstrated a spectrally-resolved single-shot wavefront sensor for broadband high-harmonic beams. 
To achieve this capability, we designed a specific Hartmann mask that incorporates transmission gratings to acquire wavefront measurements of the individual harmonics. A detailed analysis of a single measured pattern provides information on harmonic wavefronts, relative intensities, and spectral linewidths. Given sufficient sensitivity and incident flux, this information can in principle be retrieved for individual HHG pulses, rather than for an average over many shots. This feature opens up the prospect of detecting shot-to-shot variations of HHG beams, which has already been shown to be important information for sensitive experiments~\cite{Kunzel:15}.

The calibration procedure of the SHM is not more complicated than for a conventional Hartmann sensor, as only the SHM-camera distance and the orientation of the mask need to be determined, and we have shown that these parameters can even be achieved from a wavefront measurement in a self-consistent manner, if several harmonics are present in the beam. 
Although the current work demonstrates spectrally resolved wavefront measurements for wavelengths between 25 and 50~nm, this technique is not limited to these wavelengths. By changing the SHM design parameters such as the grating pitch and distance to the camera, the SHM method can be extended to wavelengths ranging from soft x-rays to infrared. By further refining the analysis procedure, it may be possible to extend the SHM technology to the characterization of more complex or partially continuous spectra.

\section*{Funding Information}
The project has received funding from the European Research Council (ERC) (ERC-StG 637476) and the Netherlands Organisation for Scientific Research (NWO).

\bibliographystyle{osajnl}
\bibliography{REF/library}

\end{document}